\documentclass[11pt]{elsart3}

\usepackage{graphicx}
\usepackage{natbib}
\usepackage{amssymb}

\newcommand\arcdeg{\mbox{$^\circ$}}%
\newcommand\arcmin{\mbox{$^\prime$}}%
\newcommand\arcsec{\mbox{$^{\prime\prime}$}}%
\newcommand{\degree}{$^\circ$}
\newcommand\sun{\odot}%

\newcommand\apjl{{Astrophys. J.~}}
\newcommand\araa{{Annu. Rev. Astron. Astrophys.~}}
\newcommand\apj{{Astrophys. J.~}} 
\newcommand\apjs{{Astrophys. J. Suppl. Ser. ~}} 
 
\newcommand\aap{{Astron. Astrophys.~}}

\newcommand\nodata{ ~$\cdots$~ }

\begin{document}
  \begin{frontmatter}

\title{Organic Molecules in the Hot Corinos and Circumstellar Disks
       of IRAS 16293-2422}

\author[NTNU]{Hui-Chun Huang\corauthref{cor}},\corauth[cor]{Corresponding
	author.} 
\ead{hspring@sgrb2.geos.ntnu.edu.tw} 
\author[NTNU]{Yi-Jehng Kuan \thanksref{now}}, \thanks[now]{Also at: Academia
Sinica Institute of Astronomy \& Astrophysics (ASIAA), Taipei 106, Taiwan, ROC}
\author[Ames]{Steven B. Charnley},
\author[ASIAA]{Naomi Hirano},
\author[CfA]{Shigehisa Takakuwa},
\author[CfA]{Tyler L. Bourke}

  \address[NTNU]{Department of Earth Sciences, National Taiwan Normal 
   University, 88 Sec.4 Ting-Chou Rd., Taipei 116, Taiwan, ROC}     
  \address[Ames]{Space Science Division, MS 245-3, NASA Ames Research
   Center, Moffett Field, CA 94035, USA}
  \address[ASIAA]{Academia Sinica Institute of Astronomy \& 
   Astrophysics, P. O. Box 23-141, Taipei 106, Taiwan, ROC} 
  \address[CfA]{Harvard-Smithsonian Center for Astrophysics,
   Submillimeter Array Project, 645 N. A'ohoku Place, Hilo, HI 96721}
 

\begin{abstract}

Using the recently-commissioned Submillimeter Array (SMA), we have
detected several complex organic molecules, including (CH$_3$)$_2$O,
C$_2$H$_5$OH, C$_2$H$_5$CN, and tentatively CH$_2$CDCN, toward the
protostellar
hot cores of IRAS 16293-2422 at arcsecond-resolution ($\lesssim$ 400
AU in radius). Vibrationally excited transitions of SO, SO$_2$ and HCN
with energy levels up to 1800 K were also observed. In addition to the
other organic molecules (HC$_3$N, CH$_2$CO, CH$_3$OH, CH$_2$CHCN and
HCOOCH$_3$) previously reported by us \citep{kua04}, these results
clearly indicate the existence of a rich organic chemistry in low-mass
`hot corinos'. From the observation of optically thin HC$^{15}$N 
emission, we conclude I16293A is a rotating circumstellar disk lying
along the north-south direction $\sim$10\degree~to the east and with
an inclination $\sim$30\degree~to the sky. We suggest that the observed
vibrational SO and SO$_2$ emission may originate from shock waves near
or in the circumstellar disks. Between the two cores, we find a strong
anticorrelation in emission from C$_2$H$_5$OH and C$_2$H$_5$CN. The
relative contribution of gas phase and grain-surface chemistries to the
production of the observed complex molecules is discussed. We point out
the shortcomings underlying recent claims that all the O-bearing
organics are formed on grains.

The presence of so many well-known interstellar molecules in {\it
solar-type hot corinos} strengthens the link between molecular cloud
chemistry, the starting materials of protoplanetary disks such as the
protosolar nebula, and the composition of comets. Establishing the fine
details of this connection is crucial in answering fundamental questions
concerning the importance of galactic astrochemistry for astrobiology.

\end{abstract}

 
\end{frontmatter} 

 

\section{Introduction}

Bombardment of the early Earth by comets and asteroids delivered large
amounts of organic materials (Chyba et al. 1990). This mechanism could
be of general importance for astrobiology since it is likely that
terrestrial planets in extrasolar nebulae experienced similar impacts.
Hence, the products of extraterrestrial organic chemistry could, in
principle, allow a prebiotic chemistry, similar to which occurred
on Earth, to begin throughout the Galaxy. The dominant molecular
composition of cometary ices is quite similar to that observed in the
gas and ices around {\it massive} protostars (Ehrenfreund \& Charnley
2000). However, to accurately make a connection between interstellar
organics and the chemicals initially available to the comet-forming
disks around Sun-like stars, high angular-resolution spectral
observations of {\it low-mass} protostars are necessary. 

The IRAS 16293-2422 core (hereafter I16293) is a young (Class 0)
region of low-mass star formation. It is located 160 pc from the Sun
in the $\rho$~Ophiuchus cloud complex. \citet{wal86}, from observed
line profiles, first identified cloud collapse in I16293, and later
\citet{zho95} modeled the I16293 system as an in-fall cloud with
rotation. The presence of a double bipolar outflow system around I16293
indicates ongoing mass accretion \citep{wal88,miz90,cas01,hir01,lis02,sta04}.
In fact, interferometric observations have shown that I16293
is a protobinary system. The two protostars, surrounded by a dense
circumbinary envelope, are separated by $\sim$5.2\arcsec, and have been
designated I16293A and I16293B, to distinguish, respectively, the SE
and NW components \citep{woo89,mun92,wal93}. From recent subarcsecond
3-mm continuum observations, \citet{loo00} derived a mass of 0.61
M$_{\sun}$ for I16293B, and a slightly lower one for I16293A
(0.49 M$_{\sun}$).

Millimeter and submillimeter line surveys using single-dish telescopes
have found many organic molecules in I16293, including CH$_3$OH, CH$_3$CN,
HC$_3$N and H$_2$CO, and suggested the existence of an inner 
($\lesssim$10\arcsec) hot region of dense gas \citep{bla94,van95}.
However, molecular species such as HCOOH, CH$_3$CHO, ${\rm CH_3OCH_3}$
and ${\rm C_2H_5CN}$, normally present in the hot molecular cores
(HMCs) associated with massive protostars, were not detected. Because
emission from the circumstellar disks was considered too dilute in their
20\arcsec~beam, \citet{van95} suggested that an interaction between the
outflow and the circumbinary envelope was responsible for the presence
of these organics. From modeling of H$_2$O data \citep{cec99} and
analysis of single-dish H$_2$CO and H$_2^{13}$CO data, \citet{cea00}
concluded that the emitting region is at around 100 K and very compact
($\sim$ 2\arcsec$-$3\arcsec), and first suggested that I16293 contained
a hot molecular core about 150 AU in size where icy grain mantles are
evaporated from the hot dust (T$_{\rm dust} \gtrsim$ 100 K).

Subsequent single-dish observations have lent further support the
{\it hot core} scenario \citep{sch02,caz03} and I16293 is now known to
have a rich molecular inventory similar to the HMCs. HMCs typically
contain high abundances of complex organic molecules, and are defined
as compact, dense and warm regions with gas densities $n_{\rm H_2}
\gtrsim 10^7~{\rm cm}^{-3}$ and $T \gtrsim 100$ K \citep{van98}. To
date, over thirty different molecular species (excluding isotopomers)
have been discovered in this low-mass protostellar source, among them
are complex organic molecules such as HCOOH, CH$_3$CN, CH$_3$OH,
CH$_3$CCH, CH$_3$CHO, HCOOCH$_3$, ${\rm CH_3OCH_3}$, ${\rm C_2H_5CN}$
and CH$_3$COOH \citep{bla94,van95,cec98,sch02,caz03}. Doubly- and
triply-deuterated molecular species such as D$_2$CO, CHD$_2$OH and
CD$_3$OH were also detected in I16293 \citep{loi00,par02,par04}.

However, all the {\it hot core} interpretations were derived from
single-dish observations which suffered from insufficient angular
resolution. The precise nature of the I16293 hot core could therefore
not be determined until very recently. By conducting arcsecond-resolution
submillimeter interferometric observations, with the detections of
complex organic molecules CH$_3$OH, CH$_2$CHCN and HCOOCH$_3$,
\citet{kua04} first uncovered the existence of two, rather than one,
compact {\it hot molecular cores} associated with each of the protobinary
components, and also evident from their thermal dust emission.
\citet{kua04} further suggested the high abundances of
organic molecules indicate that icy mantles have recently been
evaporated in the inner star-forming core. Based on millimeter
interferometric observations of CH$_3$CN and HCOOCH$_3$, \citet{bot04}
also came to the conclusion that the grain mantle evaporation scenario
is the source of complex molecules in I16293. To distinguish them from
massive HMCs, low-mass hot molecular cores, as found in I16293 and NGC
1333-IRAS4 have been referred to as {\it hot corinos} \citep{bot04}.

In this paper we report the detection and mapping of complex organic
molecules and sulfur-bearing molecules in the I16293 hot corinos.
Related observations have been previously reported by \citep{kua04} and
are important for a better understanding of the small-scale chemical
differentiation around low-mass protostars. The detection of
highly-excited transitions of simpler molecules, plus the kinematic
evidence for the existence of a possible protostellar disk, are also
presented and discussed.

\section{Observations}

Together with the close proximity of I16293, the arcsecond resolution
(1\arcsec~=160 AU) of the SMA (the Submillimeter Array\footnote{The
Submillimeter Array \citep{ho04} is a joint project between the Smithsonian
Astrophysical Observatory and the Academia Sinica Institute of Astronomy
and Astrophysics, and is funded by the Smithsonian Institution and the
Academia Sinica}), combined with high dust temperatures
($\gtrsim$~100 K) and H$_2$ density ($\sim10^7~{\rm cm}^{-3}$), makes
the SMA a perfect tool for directly imaging the higher excited submillimeter
molecular emission from the I16293 core. The SMA observations were
carried out in 2003 March (compact configuration) and July (extended
configuration) with 5 antennae, and were centered at $\alpha$(J2000) =
16$^h32^m22^s$.91, $\delta$(J2000) = -24\arcdeg28\arcmin35\arcsec.52.

The digital correlator was configured with eight chunks each of 104 MHz
bandwidth and 128 channels, except one chunk, giving a frequency
resolution of 0.812 MHz. The frequency range covered by the two sidebands
are: 343.555$-$344.225 GHz (lower sideband) and 354.211$-$354.881 GHz
(upper sideband). The synthesized beams are
$\sim$1\arcsec.3$\times$2\arcsec.7 at 344 GHz and
$\sim$1\arcsec.1$\times$2\arcsec.5 at 354 GHz (natural weighting);
these beam sizes correspond to a linear scale of $\sim$200$\times$400
AU at a distance of 160 pc. NRAO530 and 1743-038 were observed for
phase and amplitude calibration. A 25\% uncertainty of the flux scale
is estimated.


\section{Results and Discussion}

\begin{figure*}
\hspace{+2.5cm}
\includegraphics[width=4.0in]{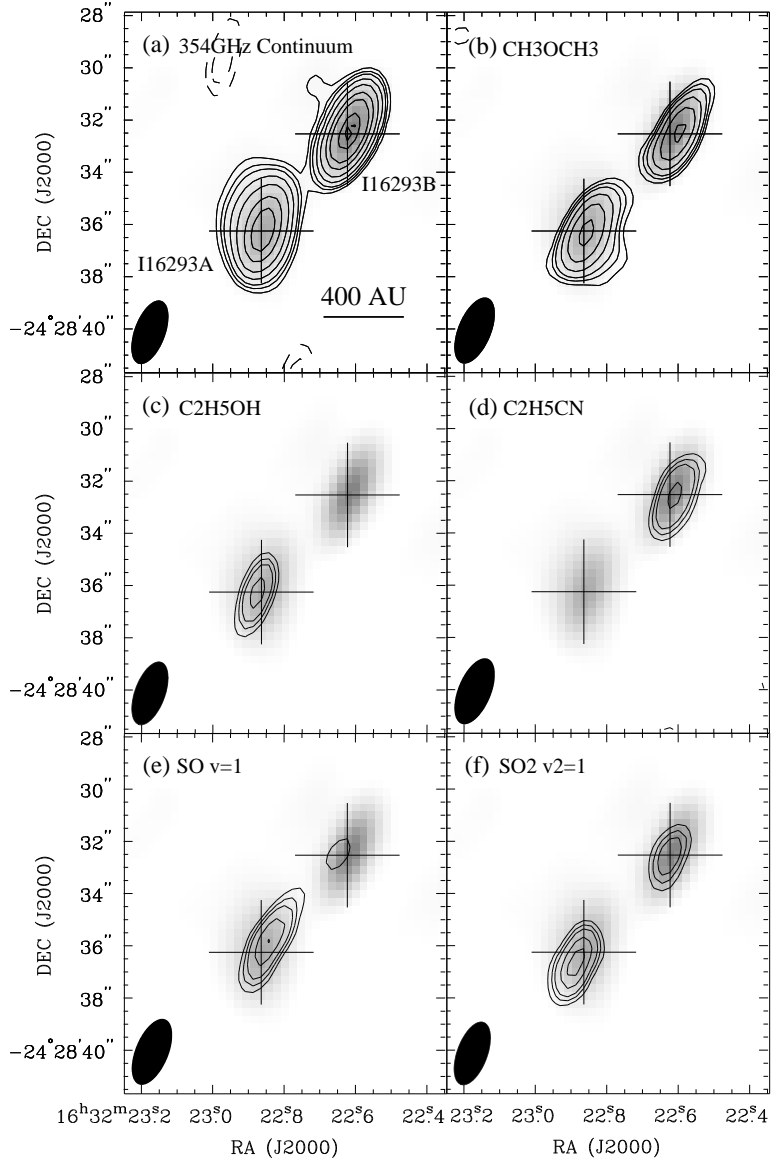}   
\caption{Spectral images of large organic molecules and two
sulfur-containing molecules toward IRAS 16293-2422.
(a) Continuum at 354 GHz; (b) spectral emission
of (CH$_3$)$_2$O; (c) C$_2$H$_5$OH emission; (d) C$_2$H$_5$CN;
vibrationally excited (e) SO {\it v} = 1 and (f) SO$_2$
{\it v$_2$} = 1 emission.
Plus signs mark the positions of I16293A and I16293B hot cores. The
angular size for a linear scale of 400 AU is shown in (a). The grey
scale denotes the 354 GHz continuum. The dark ellipse represents the
HPBW of the synthesized beam. Contours are shown at 3, 4, 5, 7
$\sigma$ levels in general, then at irregular intervals up to the
peak values; dashed lines indicate contours at negative 3 and 4
$\sigma$ levels.\label{fig1}}
\end{figure*}

Table~\ref{tbl-1} lists the molecular transitions detected toward the
I16293 hot corinos; it includes highly-excited vibrational lines of SO
($v$=1), SO$_2$ ($v_2$=1) and HCN ($v_2$=1), the molecular lines of
sulfur-containing molecules $^{34}$SO$_2$, OC$^{34}$S and H$_2$CS,
those of complex organic molecules (CH$_3$)$_2$O, $g$-C$_2$H$_5$OH
and C$_2$H$_5$CN, and the tentatively detected CH$_2$CDCN lines,
plus the optically thin HC$^{15}$N line.
Continuum emission at 354 GHz was imaged at a noise level of 0.11
Jy beam$^{-1}$ (Fig.~\ref{fig1}{\it a}). The peak position of I16293A,
with a flux density of 4.97$\pm$0.49 Jy, is at $\alpha$(J2000) =
16$^h32^m22^s$.86, $\delta$(J2000) = -24\arcdeg28\arcmin36\arcsec.25,
and of I16293B, at $\alpha$(J2000) = 16$^h32^m22^s$.62,
$\delta$(J2000) = -24\arcdeg28\arcmin32\arcsec.53 with a flux density
of 5.14$\pm$0.61 Jy, which is slightly higher than that of I16293A.


\subsection{The Hot Corinos}

\begin{figure*}
\hspace{+2.2cm}
\includegraphics{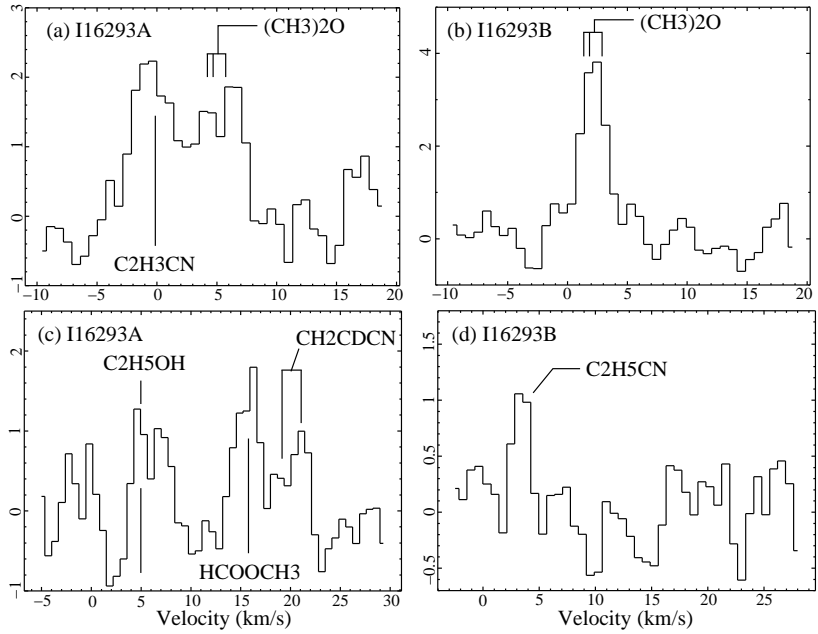}
\caption{Sample spectra of large organic molecules toward IRAS
16293-2422. The left column shows spectra taken at I16293A; the right
column, at I16293B. The (CH$_3$)$_2$O emission is shown in (a) and
(b); the C$_2$H$_5$OH line in (c); and C$_2$H$_5$CN in (d). The weaker
spectral appearance of the C$_2$H$_3$CN emission (reported in
\citep{kua04}) in panel (a) and the HCOOCH$_3$ (also reported in
\citep{kua04}) and CH$_2$CDCN lines in panel (c) is because spectra
(a) and (c) are taken, respectively, at the peak-emission positions of
(CH$_3$)$_2$O and C$_2$H$_5$OH integrated-intensity maps. The combined
effect of the intrinsic C$_2$H$_3$CN emission weakness and the near
half-beamwidth offset of C$_2$H$_3$CN peak position ($\sim1\arcsec$
southeast) from the (CH$_3$)$_2$O peak toward I16293B results in the
apparent disappearance of C$_2$H$_3$CN emission in panel (b) of the
(CH$_3$)$_2$O spectrum. All spectra were Hanning smoothed for better
S/N ratios.\label{fig2}}
\end{figure*}

Sample images of three large organic molecules are shown in
Figs.~\ref{fig1}{\it b-d}; these molecules are also commonly detected
in the HMCs of massive star-forming regions. All spectral emission, within
a beam-convolved size of $\sim$200$\times$400 AU in radius, appears to
originate from the two protobinary components I16293A and I16293B.
Given that all spectral lines shown in Figure 1 have lower-energy
levels above 100 K, it is reasonable to assume that the major spectral
emission comes from the inner dense cores, as clearly reflected in
these spectral images, and is not an accumulated effect along the
line-of-sight. In some cases, molecular emission was only detected
toward one protostellar core. A single-source detection could be due to
the combined effect of abundance and excitation differences between the
two sources. It should be noted that some of the spectral lines, which
were present at a $\sim$ 2-$\sigma$~level ($\lesssim$ 0.6 Jy beam$^{-1}$)
in the channel maps, are reported as nondetected lines in this study.
Fig.~\ref{fig2} shows sample spectra for four large organic molecules in
either I16293A or I16293B.

It is known that massive protostellar cores can be very hot. HCN
($v_2$=2) and SO$_2$ ($v_2$=1) transitions with energy levels up to
2000 K and 1400 K, respectively, were observed toward HMCs in Orion KL
and Sgr B2(M) \citep{sch97,sut91}. By observing HCN lines at 797 GHz,
\citet{boo01} derived $T_{\rm ex} =$ 720 K in the massive protostar GL 2591.
However, in the case of low-mass protostars like I16293, where the total
bolometric luminosity is only about 30-40 $L_{\sun}$ \citep{mun92}, it
was thought that the surrounding dust could not be heated sufficiently
to create a hot core environment, except in the innermost 100 AU
\citep{van98}. The detection of highly excited transitions, such as HCN
($v_2$=1) 4$-$3, SO ($v$=1) 8$_9$$-$7$_8$ (Fig. 1(e)) and SO$_2$
($v_2$=1) (Fig. 1(f)) 46$_{5,41}$$-$46$_{4,42}$, with their respective
lower energy levels occurring 1050 K, 1660 K and 1800 K above ground,
clearly demonstrates that the innermost regions of the two protostellar
cores are very hot.
The vibrationally excited transitions are likely to be maintained by
far-infrared pumping. These high-resolution spectral images clearly
provide further strong observational evidence for the existence of
low-mass hot cores similar to the HMCs found in massive star-forming
regions.

\subsection{I16293A: A Circumstellar Disk?}

\begin{figure*}
\includegraphics[width=6.0in]{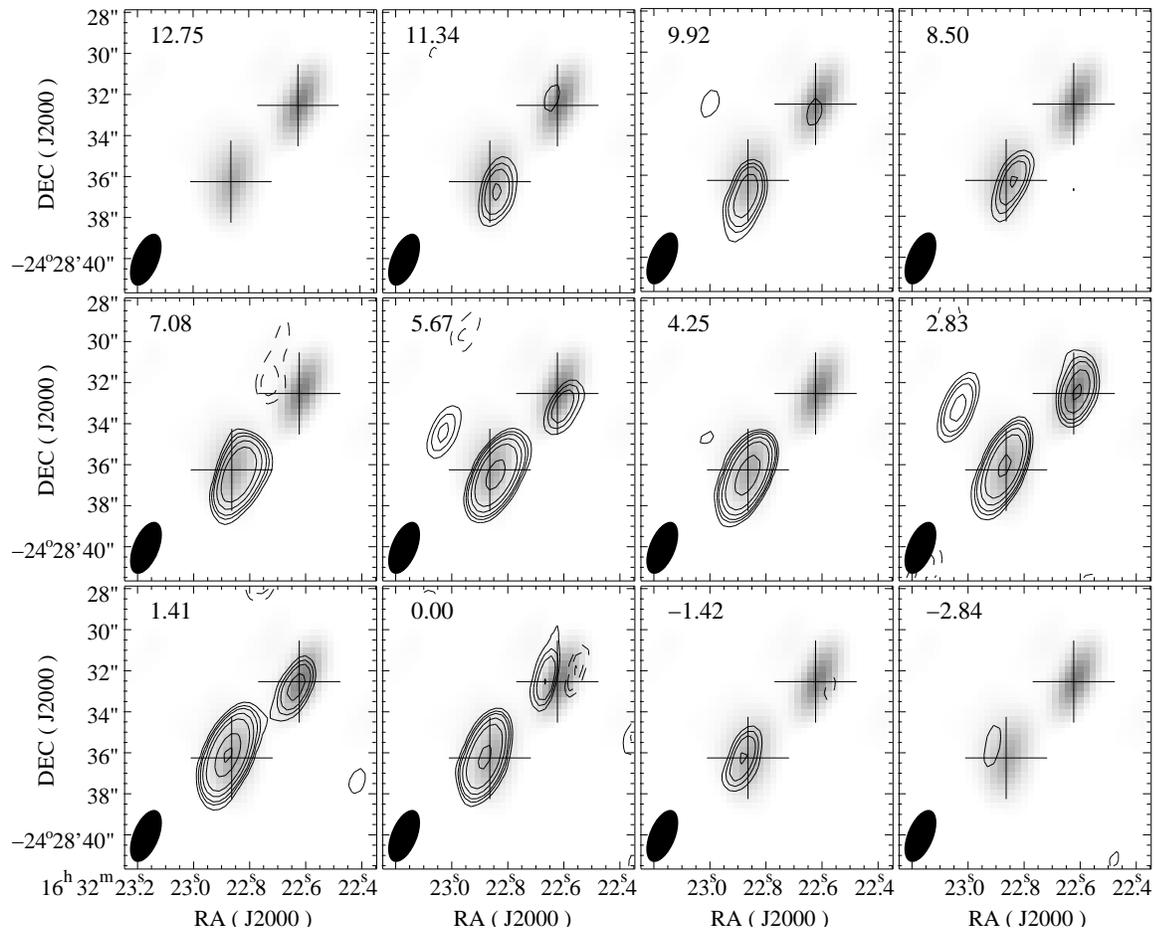}
\caption{Velocity channel maps of HC$^{15}$N 4-3 emission toward I16293A.
 The same figure labels and symbol conventions used in Fig. 1 are adopted
 here. The number at the upper-left corner of each panel is the LSR velocity
 of that particular channel.\label{fig3}}
\end{figure*}

I16293 is a region full of dynamic activity, such as double bipolar
outflows, a surrounding in-fall cloud and a circumbinary envelope,
which makes an unambiguous kinematic interpretation difficult. Thus, to
kinematically separate the inner hot cores from other gas motions in
the region, observations of optically thin molecular emission of
a high-density tracer would be ideal. HC$^{15}$N is a rare isotope of
the high-density tracer HCN, thus emission of HC$^{15}$N is most likely
to be optically thin which would trace the densest regions in I16293
and, in principle, better avoid any potential contamination from the
ambient cloud and/or gas outflows. By examining the velocity channel
maps of the HC$^{15}$N J=4-3 emission (Fig.~\ref{fig3}), molecular gas
motion is apparent. This can be seen in moving across the center
position of I16293A, from the direction south-southwest at higher
velocities (V$_{\rm LSR}$ $\sim$ 11 to 5 km s$^{-1}$) to the 
north-northeast at lower velocities (V$_{\rm LSR}$ $\sim$ 3 to -3
km s$^{-1}$). Likewise, OC$^{34}$S and HCN ($v_2$=1) channel maps
(not shown here) also show a velocity gradient toward I16293A at a
position angle (PA, measured from the north to the east
counterclockwise) between 10\degree~and 30\degree, similar to that
of HC$^{15}$N.

\begin{figure*}
\hspace{2.0cm}
\includegraphics[width=4.4in]{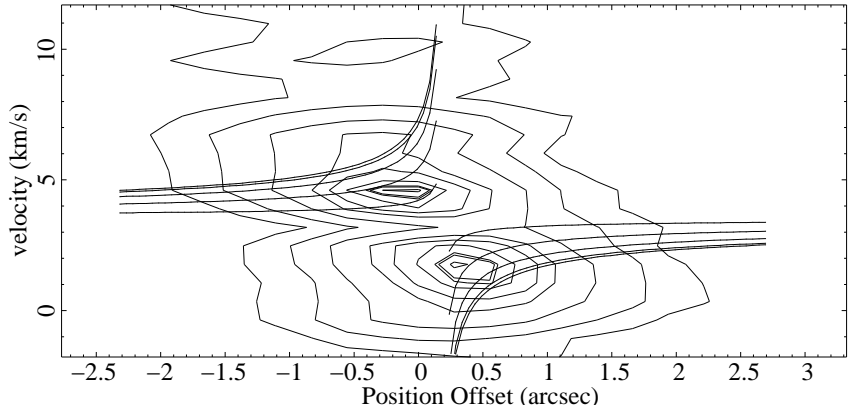}
\caption{HC$^{15}$N position-velocity diagram at I16293A. The velocity
	 cut is taken at the I16293A position (offset position =
	 +0.2\arcsec) at a position angle PA = 10\degree. Contours
	 denote the HC$^{15}$N emission at 3-$\sigma$ level and above.
	 The thick solid curves represent the velocity fields derived
	 from a rotating keplerian disk model at various inclinations:
	 $i$ = 10\degree~(the innermost one), 30\degree, 50\degree,
	 70\degree~and 90\degree~(the outermost one). A central mass
	 of 0.49 M$_{\sun}$ and a systemic velocity of 3.7 km s$^{-1}$
	 are used for the modeling.\label{fig4}}
\end{figure*}

A position-velocity cut centered at I16293A and a PA = 10\degree~
is shown in Fig.~\ref{fig4}. Two distinguished velocity components,
with their peak-emission positions separated by $\sim$3.0 km s$^{-1}$
in velocity and $\sim$0.4\arcsec~($\sim$60 AU) in position, are
immediately visible. By modeling the kinematic structure of I16293A thus
revealed, we find it can be fitted nicely by a rotating keplerian disk
inclined $\sim$30\degree~with respect to the sky (an inclination
angle $i$ = $\sim$30\degree, Fig.~\ref{fig4}); here a mass of 0.49
M$_{\sun}$ for the central protostar \citep{loo00} and a systemic
velocity of 3.7 km s$^{-1}$ are adopted for the model. Note that in
the channel maps (Fig.~\ref{fig3}), the orientation of the unresolved
I16293A image of HC$^{15}$N emission is largely a result of beam
convolution, and is not the true orientation of the disk which is
tilted $\sim$10\degree~east of the north. The I16293A
circumstellar disk is distinct from the circumbinary disk previously
reported in the literature.

Our finding is consistent with the subarcsecond continuum images of
I16293A \citep{mun92,loo00} which also show a north-south elongated
structure off slightly to the east. This rotating disk scenario
is further supported by more recent observations of the double bipolar
outflows emanating from the I16293 star-forming core \citep{lis02,sta04};
the ``E-W outflow'' of the two outflows is in fact in the direction
perpendicular to the I16293A circumstellar disk. Since the two HC$^{15}$N
emission peaks in the position-velocity diagram is only 0.4\arcsec~or
60 AU apart, it is reasonable to believe that most of the molecular gas
in the disk resides at a distance merely 30 AU from the central
protostar. This raises the distinct possibility that the rotating disk of
I16293A is really a protoplanetary disk.

\subsection{Chemical Composition}

Table~\ref{tbl-2} lists the beam-averaged molecule column densities
($N_{\rm A}$ and $N_{\rm B}$) and derived fractional abundances (i.e.
$X_{\rm A}=N_{\rm A}/N({{\rm H}_2})$) in both hot corinos. For
completeness, Table~\ref{tbl-3} lists the column densities and fractional
abundances of those molecules reported by \citep{kua04}. Column densities
were derived assuming optically thin lines in local thermodynamic
equilibrium and were measured at the peak emission positions of each
integrated spectral line. This approach is sufficient only for
obtaining first-order estimates since excitation temperatures for
individual molecules could differ, and some lines may have non-trivial
optical depths. Because of the very high gas density of $n_{\rm H_2}
\gtrsim 10^7$ \citep{van95,cea00,ceb00}, gas and dust are expected to be
thermally well-coupled so an excitation temperature $T_{\rm ex}$
$\simeq$~$T_{\rm dust}$ = 100 K was adopted \citep{kua04}, except for
molecular transitions with energy levels above 500 cm$^{-1}$
($\gtrsim$~700 K) where $T_{\rm ex}$ = 300 K was used \citep{sch02}.
In addition, we assume the vibrational excitation temperature is similar
to the rotational excitation temperature with $T_{\rm ex}$(vib)
$\simeq$~$T_{\rm ex}$(rot).
Here a gas column density $N({\rm H_2}) = 1.6 \times 10^{24}$ cm$^{-2}$
\citep{sch02} was adopted for both corinos. The derived
fractional abundances and relative column densities are in good
agreement with the values observed within a similar frequency range
(330-355 GHz) for the Orion KL and Sgr B2(M) massive HMCs
\citep{sut91,sut95,sch97}. For the rare isotopes detected, we find
SO$_2$/$^{34}$SO$_2$ $\simeq$ 10 and HCN(GND)/HC$^{15}$N $\simeq$ 
8$-$29, also in general agreement with HMC values \citep{num00,sch97}.

When compared to some previous single-dish observations toward I16293
\citep{bla94,van95,caz03}, many molecular abundances derived using the
smaller SMA beam (SO, SO$_2$, H$_2$CS and C$_2$H$_5$CN) are higher in
general. This indicates that the observed molecular emission is mainly
from the compact cores (or disks), and explains why SO, SO$_2$ and
H$_2$CS abundances derived from modeling higher energy transitions
\citep{sch02} in the {\it inner} I16293 core show good agreement with
our values. The ground 4$-$3 HCN emission toward I16293 is extended
and diffuse \citep{kua04}. The fact that the abundance derived from
the vibrational bending state HCN is an order of magnitude higher than
the ground HCN toward I16293A suggests more HCN resides in the compact
protostellar disk of I16293A.


The high abundances of organic molecules found in both hot corinos can
be best explained by icy mantle evaporation \citep{cha92}. The
abundances derived by us for most of those molecules observed in both
sources are within a factor of two of each other, as to be expected
if both cores collapsed from the same cloud material. There are,
however, marked compositional differences between I16293A and I16293B.
The nondetection of SO emission toward I16293B cannot be due to
excitation; it is more likely due to an actual chemical difference
since an even higher excitation line of SO$_2$ is seen in I16293B.
Similarly, \citet{mun92} did not detect SO in I16293B though they also
observed it in I16293A. Although the vibrationally excited lines of SO
and SO$_2$ can still be populated by an excitation temperature of
$\sim$100 K, it is probable that these molecular lines actually arise
well within the inner hot cores, perhaps in circumstellar disks.
\citet{wal94} previously concluded that vibrationally excited CS
emission in I16293 emanated from gas with $T \gtrsim$~1000 K, and at
densities such that the only likely environment was in shocks within
the protostellar disk. \citet{van98} also suggested a possible existence
of shocked zones in the innermost 100 AU. Complete conversion of SO to
SO$_2$ in shock waves could explain the lack of SO in I16293B.

Apart from the striking anticorrelations in the presence of C$_2$H$_5$CN
and C$_2$H$_5$OH between I16293A and I16293B, there appears to be no
dichotomy between N-containing and O-containing organic molecules
similar to that found in massive star-forming regions \citep{wyr97}.
However, these anticorrelations
could pose problems for grain chemistry scheme based on H atom additions.
Together with organic molecules reported previously in \citep{kua04},
these posit that HC$_3$N and CH$_2$CHCN are precursors of C$_2$H$_5$CN,
and that CH$_2$CO is a possible precursor of C$_2$H$_5$OH
\citep{cha92,cha04}. The HC$_3$N/CH$_2$CHCN ratios in both sources are
similar but C$_2$H$_5$CN is only detected in I16293B, and at a large
abundance. Ethanol is only detected in I16293A but ketene, its putative
precursor in some grain-surface reaction schemes \citep{cha97}, is not.
Low-level emission from CH$_2$CO and C$_2$H$_5$CN was seen in channel maps
of I16293A, as was C$_2$H$_5$OH emission in I16293B. Nevertheless the
observations suggest that these molecules are chemically differentiated
between the two sources.
This apparent anticorrelation between precursors and end
products in I16293A and I16293B, as well as the relative abundances of the
species involved in the surface chemistry, may eventually allow us to
constrain theories of grain surface chemistry. For example, reduction of
ketene to ethanol via acetaldehyde ($\rm CH_3CHO$) is is one possible
pathway (Charnley 1997). Alternatively, reduction and oxidation of
acetylene can produce both acetaldehyde and ethanol, as well as ethylene
oxide and vinyl alcohol (Charnley 2004). Our observations suggest that 
the latter pathway may have dominated in the prestellar phase of I16293
and hence that a similar differentiation in the $\rm CH_3CHO$ and
$c$-C$_2$H$_4$CO abundances should exist between I16293A and I16293B.
More observations, as well as better excitation and abundance
determinations are necessary to pursue this avenue of research.

Indeed, a fundamental issue is to determine which of the organic
molecules found in HMCs and hot corinos actually do originate in
grain-surface chemistry, or whether post-evaporative production in
the hot gas is necessary (Charnley et al. 1992).
Ion-molecule production of methyl formate, $\rm HCOOCH_3 $, and of
dimethyl ether, $\rm (CH_3)_2O$, has been proposed and involves,
respectively, methyl cation transfer from protonated methanol to
formaldehyde and methanol \citep{bla87}.
When the methanol and formaldehyde molecules are injected from grain
surfaces, these reaction pathways can reproduce the $\rm (CH_3)_2O$
and $\rm HCOOCH_3 $ abundances observed in HMCs \citep{cha92}.
However, single-dish observations of the I16293 core led Cazaux et al.
(2003) to proclaim that all organics were produced on grain surfaces.
The abundances of O-containing molecules derived by Cazaux et al.
(2003) were almost all much higher than found elsewhere and this, when
compared to the methanol abundance measured by \citet{sch02}, led these
authors to conclude that all the O-containing molecules, particularly
methyl formate and dimethyl ether, were formed on grains.

It is extremely dangerous to make this generalization for several
reasons. First, Cazaux et al. (2003) argued that, because the observed
$\rm HCOOCH_3 $/$\rm CH_3OH $ is close to unity in I16293, the above
gas phase production could be ruled out. In the case of $\rm HCOOCH_3 $,
ion-molecule experiments {\it do show} that reaction of $\rm CH_3OH_2^+ $
with $\rm H_2CO $ does not produce protonated methyl formate, and so an
origin in grain-surface reactions or other ion-molecule reactions,
perhaps involving HCOOH, is more likely \citep{cha97}. On the other hand,
it has been experimentally verified that gas phase self-methylation of
methanol will lead to $\rm (CH_3)_2O$. A second, and more serious, problem
lies in the fact that Cazaux et al. constructed their
$\rm HCOOCH_3 $/$\rm CH_3OH $ ratio using abundances from two different
observations. Fractional abundances derived by \citet{sch02} were derived,
as here, using a gas column density of $N({\rm H_2}) =
1.6 \times 10^{24}$ cm$^{-2}$, whereas Cazaux et al. (2003) used
$N({\rm H_2}) = 7.5 \times 10^{22}$ cm$^{-2}$. This explains why the
abundances derived by Cazaux et al. (2003) are so high. Third,
interferometric observations of $\rm HCOOCH_3 $ in both I16293A and
I16293B show that its emission comes from the two hot corinos. Abundances
derived using more realistic (i.e. larger) values of $N({\rm H_2})
(\gtrsim 10^{24}$ cm$^{-2}$) lead to $\rm HCOOCH_3 $ and $\rm (CH_3)_2O$
abundances ($\sim 10^{-9}-10^{-8}$), and ratios relative to $\rm CH_3OH $
($\sim 0.01-0.1$), that are more in accord with those found elsewhere
(this work; \citet{kua04,bot04}).
Based on interferometry of $\rm HCOOCH_3 $ and $\rm CH_3CN$ emission in
both the I16293 hot corinos, \citet{bot04} also concluded that all the
organic molecules were evaporated products of grain-surface chemistry.
While a gas phase origin for $\rm HCOOCH_3 $ cannot yet be ruled out,
$\rm CH_3CN$, like $\rm (CH_3)_2O$, can be formed easily in hot gas
\citep{rod01}.

In conclusion, the issues raised above show that $\rm HCOOCH_3 $ is 
{\it not} a good molecule to discriminate between a global gas-phase
or grain-surface origin for all other molecules. Nevertheless, it is 
almost certain that some molecules, particularly $\rm (CH_3)_2O$, are 
produced through ion-molecule reactions. Similarly, it is also quite 
certain that some organic compounds seen in I16293, those that do not 
appear to have a viable gas-phase production chemistry, are formed 
solely on grains: HCOOH, $\rm CH_3OH$, $\rm C_2H_5OH$, HNCO, $\rm CH_2CO$, 
$\rm CH_3CHO$ \citep{cha01}. 

It is important to try and estimate the relative evolutionary state of
the I16293 sources, as this may provide clues to the chemistry that
occurred just prior to the formation of the protosolar nebula.
\citet{sta04} have suggested that, whereas I16293A is a Class 0 object,
I16293B is a T Tauri star and consequently more evolved.
A large degree of deuterium fractionation is observed toward I16293
\citep{loi00,par02,par04} and the CH$_2$CHCN/CH$_2$CDCN ratios found
here ($\sim$3 in I16293A and $\sim$4 in I16293B) suggest both sources
are in an early phase of chemical evolution. It is evident that the
spectral-peak and integrated-line intensities are in general stronger
toward I16293A which may imply I16293A is warmer and hence the more
evolved of the two. $c$-C$_3$H$_2$, CH$_2$CO and C$_2$H$_5$CN are only
detected in I16293B and, as their fractional abundances are higher than
in the more evolved HMCs (this work; \citet{kua04}), this may also be
a further indication that I16293B is the less evolved of the two.
On the other hand, gas-phase chemical evolution driven by prompt 
evaporation of mantle molecules could allow daughter/parent abundance
ratios to be used for estimating the relative post-evaporative ages of
HMCs \citep{cha92}. The observed $\rm (CH_3)_2O$/$\rm CH_3OH$ ratios
in I16293A (0.006) and I16293B (0.016) appear to indicate that, in fact,
I16293B is the more evolved; assuming that the higher methanol abundance
of I16293A is representative of that originally injected into both hot
corinos. Clearly, given the caveats surrounding the abundance
determinations, it may be premature to make definitive estimates of
chemical age until more secure values are available.


\section{Conclusions}

Together with previous high-resolution SMA studies, a rich organic
inventory of large organic molecules such as HC$_3$N, CH$_2$CO,
CH$_3$OH, CH$_2$CHCN, HCOOCH$_3$, (CH$_3$)$_2$O, C$_2$H$_5$OH and
C$_2$H$_5$CN is revealed in low-mass hot corinos. By observing the
optically thin HC$^{15}$N emission, we find I16293A is in fact a
rotating circumstellar disk distinct from the circumbinary disk,
lying along the north-south direction with a position angle
$\sim$10\degree~to the east, and is inclined $\sim$30\degree~to the
sky. Hence there is potentially a strong link between interstellar
organics and prebiotic chemistry in protostellar disks \citep{ehr00}.
There are several chemical similarites between the compositions of
I16293A and I16293B, however, C$_2$H$_5$CN, CH$_2$CO and C$_2$H$_5$OH
appear to be strongly differentiated between the cores. The vibrational
excitation observed in SO and SO$_2$ suggests an active shock chemistry
in protostellar disks. Star-forming cores are formed from cold parent
molecular clouds, and later evolve into protostellar disks. Our
observations strengthen the chemical connection between dark
clouds, massive star-forming regions and solar-type hot corinos. An
important missing piece of this picture is the understanding of how
interstellar organic material can be modified and incorporated into
protostellar disks similar to our protosolar nebula, and survive
throughout the planet formation stage and beyond. Future SMA
observations of large organics in I16293 and similar cores will be
important to the study of chemical evolution of protoplanetary disks.


\section{Acknowledgements} 

The authors would like to thank the referees for the useful comments.
We are also grateful to the JPL Molecular Spectroscopy web services
(http://spec.jpl.nasa.gov/) and the Cologne Database for Molecular
Spectroscopy (CDMS, http://www.ph1.uni-koeln.de/vorhersagen/)
for making molecular laboratory data available.
The research of Y.-J. K. was supported by NSC 93-2112-M-003-003 grant.
This work was supported by NASA's Exobiology and Origins of Solar Systems
Programs through funds allocated by NASA Ames under Interchange No.
NCC2-1412 to the SETI Institute.


\clearpage

\begin{table*}
\caption{Molecular transitions detected toward IRAS 16293-2422
         hot cores.\label{tbl-1}}
\begin{center}
\begin{tabular}{llcrrr}
\noalign{\smallskip}
\hline\hline
\noalign{\smallskip}
Molecule~~~~~~~~ &  Transition & Rest Frequency  &  $E_{\rm low~}$  & 
	~~~I$_{\nu,{\rm I16293A}}$$^{\mathrm a}$~ &
 	~~~I$_{\nu,{\rm I16293B}}$$^{\mathrm a}$~ \\
 &  & $\rm (MHz)$ & (cm$^{-1}$) & (Jy bm$^{-1}$) & (Jy bm$^{-1}$) \\   
\hline
    SO ($v$=1) & 8$_9$-7$_8$ & 343829.4$^{\mathrm b}$~ & 1154.4 & 1.18$\pm$0.32 &
        \nodata \\
SO$_2$ ($v_2$=1) & 46$_{5,41}$-46$_{4,42}$ & 354624.2 & 1251.2 &
        1.59$\pm$0.20 & 1.46$\pm$0.20 \\
HCN ($v_2$=1) & 4-3~~$l$=1e & 354460.4 & 729.7 & 1.97$\pm$0.20 &
        \nodata \\
HC$^{15}$N$^{\mathrm c}$ & 4-3 & 344200.3 & 17.2 & 5.26$\pm$0.24 &
        1.72$\pm$0.24 \\
$^{34}$SO$_2$ & 19$_{8,12}$-20$_{7,13}$ & 354397.8 & 214.5 &
        1.77$\pm$0.28 & \nodata \\
OC$^{34}$S & 29-28 & 343983.3 & 160.7 & 2.58$\pm$0.31 &
        2.56$\pm$0.31 \\
H$_2$CS & 10$_{2,8}$-9$_{2,7}$ & 343810.8 & 88.2 & 1.42$\pm$0.32 &
	1.96$\pm$0.32 \\
CH$_2$CDCN & 38$_{12,26}$-37$_{12,25}$ & 354737.4$^{\mathrm d}$ &
	388.7 & 1.57$\pm$0.19 & 1.11$\pm$0.19 \\
& 38$_{12,27}$-37$_{12,26}$ & 354737.4$^{\mathrm d}$ & 388.7 & & \\
& 38$_{11,27}$-37$_{11,26}$ & 354739.2$^{\mathrm d}$ & 361.7 & & \\
& 38$_{11,28}$-37$_{11,27}$ & 354739.2$^{\mathrm d}$ & 361.7 & & \\
(CH$_3$)$_2$O & 17$_{2,16}$-16$_{1,15}$ EA & 343753.3 & 88.4 &
	1.85$\pm$0.26 & 3.79$\pm$0.26 \\
        & 17$_{2,16}$-16$_{1,15}$ AE & 343753.3 &  88.4 & & \\
        & 17$_{2,16}$-16$_{1,15}$ EE & 343754.2 &  88.4 & & \\
        & 17$_{2,16}$-16$_{1,15}$ AA & 343755.1 &  88.4 & & \\
$g$-C$_2$H$_5$OH & 20$_{3,17}$-19$_{3,16}$ & 354757.4 & 161.1 &
        1.28$\pm$0.19 & \nodata \\
C$_2$H$_5$CN & 40$_{8,32}$-40$_{7,33}$ & 343888.0 & 283.1 &
	\nodata & 1.07$\pm$0.15 \\
      \noalign{\smallskip}   
      \hline
\end{tabular}  \end{center}
\footnotesize
  \begin{flushleft}
 \footnotesize
 {$^{\mathrm a}$}{~The line intensity measured in Hanning-smoothed
  spectrum at the peak-emission position of an  integrated intensity
  map. }\\
 {$^{\mathrm b}$}{~Taken from the Cologne Database for Molecular
  Spectroscopy (CDMS, http://www.ph1.uni-koeln.de/vorhersagen/).} \\
 {$^{\mathrm c}$}{~The HC$^{15}$N data are taken from \citet{kua04}.} \\
 {$^{\mathrm d}$}{~The 354737 and 354739 MHz doublets are blended
  into a single ``line" at $\sim$354738 MHz.} 
  \end{flushleft} 
\end{table*}

\clearpage

\begin{table*}
\caption{Molecular column densities and fractional abundances in
     IRAS 16293-2422 A \& B.\label{tbl-2}}
\begin{center}
\begin{tabular}{lccccccccc}
\noalign{\smallskip}
\hline\hline
\noalign{\smallskip}
 &     &   I16293A      &    & ~ &   &  I16293B       &   &   &  \\
\cline{2-4}  \cline{6-8}
 ~Molecule~ &  $\int$ I$_{\nu} dV$$^{\mathrm a}$   &
     ~~$N_{\rm A}$$^{\mathrm b}$~~~   &  $X_{\rm A}$  &  &
     $\int$ I$_{\nu} dV$$^{\mathrm a}$  & ~~$N_{\rm B}$$^{\mathrm b}$~~~  &
       $X_{\rm B}$  & ~~~$X_{\rm HMC}$~~~  &  $X_{\rm I16293}$   \\
 &     &   (cm$^{-2}$)  &    & &   &  (cm$^{-2}$)   &   &   &  \\   
\hline
SO     & 2.69   & 4.2(16)$^{\mathrm c}$ & 2.6(-8)  & & \nodata    & \nodata &
   \nodata      & 1.9(-7)$^{\mathrm d}$ &
   3.9(-9)$^{\mathrm e}$, 2.5(-7)$^{\mathrm f}$ \\
SO$_2$ & 4.42   & 2.2(17)$^{\mathrm c}$ & 1.4(-7)  & & 3.00       &
   1.5(17)$^{\mathrm c}$   & 9.2(-8)    & 1.2(-7)$^{\mathrm d}$   &
   1.5(-9)$^{\mathrm e}$, 1.0(-7)$^{\mathrm f}$                   \\
HCN ($v_2$=1)   & 9.89     & 2.7(15)$^{\mathrm c}$ & 1.7(-9) &  & \nodata &
   \nodata      & \nodata  & 3.2(-9)$^{\mathrm g}$ & 1.9(-9)$^{\mathrm h}$ \\
$^{34}$SO$_2$   &   2.50   & 2.3(16)    & 1.4(-8)  & &
   \nodata      &  \nodata &  \nodata   &  \nodata & \nodata      \\
OC$^{34}$S      & 12.36    & 4.7(15)    & 3.0(-9)  & & 8.32       &
   3.2(15)      & 2.0(-9)  & \nodata    & \nodata                 \\
H$_2$CS & 5.94  & 1.5(15)  & 9.4(-10)   & & 5.31   & 1.3(15)      & 8.4(-10) &
   8.(-10)$^{\mathrm d}$   & 1.7(-10)$^{\mathrm e}$, 5.5(-9)$^{\mathrm f}$ \\
CH$_2$CDCN      & 3.49     & 5.0(15)    & 3.1(-9)  & & 1.83       &
   9.6(14)      & 6.0(-10) & \nodata    & \nodata  \\
(CH$_3$)$_2$O   & 8.49     & 6.6(15)    & 4.1(-9)  & & 10.32      & 8.0(15)  &
   5.0(-9)      & 8.(-9)$^{\mathrm d}$  &  2.4(-7)$^{\mathrm i}$        \\

C$_2$H$_5$OH    & 3.61     & 1.5(16)    & 9.5(-9)  & & \nodata    &
   \nodata      & \nodata  & 7.(-10)$^{\mathrm d}$ & \nodata      \\
C$_2$H$_5$CN    & \nodata  & \nodata    & \nodata  & & 2.03       & 4.3(16)  &
   2.7(-8)      & 3.(-9)$^{\mathrm d}$  & 1.2(-8)$^{\mathrm i}$   \\
\noalign{\smallskip}   
  \hline
\end{tabular}  \end{center}
\footnotesize
  \begin{flushleft}
 \footnotesize     
 $^{\mathrm a}${~In Jy bm$^{-1}$~km s$^{-1}$.}   \\
 $^{\mathrm b}${~T$_{\rm ex}$ = 100 K was assumed for all lines
            except where noted; $a(b) = a \times 10^b$.}  \\
 $^{\mathrm c}${~For transitions with energy levels above 500
            cm$^{-1}$, T$_{\rm ex}$ = 300 K is adopted.}  \\
 $^{\mathrm d}${~Orion KL hot core; from \citet{sut95}.} \\
 $^{\mathrm e}${~Single-dish observations with a $\sim20\arcsec$~beam;
   	    from \citet{bla94}.} \\
$^{\mathrm f}$ {~I16293 hot core; from models of \citet{sch02}.}    \\
$^{\mathrm g}${~Sgr B2(M) hot core; from \citet{sut91}.} \\
$^{\mathrm h}${~Single-dish observations with a
$\sim20\arcsec$~beam;
	   from \citet{van95}.} \\
$^{\mathrm i}${~Single-dish observations with
   	   $\sim10\arcsec-30\arcsec$ beam; from \citet{caz03}.}           
  \end{flushleft} 
\end{table*}

\clearpage

\begin{table}
\caption{Column densities and fractional abundances of other organic
         molecules in IRAS 16293-2422 A \& B$^{\mathrm a}$.\label{tbl-3}}
\begin{center}
\begin{tabular}{lcccc}
\noalign{\smallskip}
\hline\hline
\noalign{\smallskip}
 Molecule  & $N_{\rm A}$     & $X_{\rm A}$ & $N_{\rm B}$    & $X_{\rm B}$ \\
           &  ~~(cm$^{-2}$)~ &             & ~~(cm$^{-2}$)~ &             \\   
\hline
HCN (GND)       & 3.1(14) & 2.0(-10) & 1.7(14) & 1.0(-10) \\
HC$^{15}$N      & 1.2(14) & 7.4(-11) & 1.9(13) & 1.2(-11) \\   	 
$c$-C$_3$H$_2$  & \nodata & \nodata  & 7.2(15) & 4.5(-9)  \\
CH$_2$CO        & \nodata & \nodata  & 1.9(15) & 1.2(-9)  \\
HC$_3$N         & 6.7(14) & 4.2(-10) & 1.3(14) & 8.4(-11) \\
CH$_3$OH        & 1.1(18) & 6.8(-7)  & 5.0(17) & 3.1(-7)  \\
$^{13}$CH$_3$OH & 8.1(16) & 5.0(-8)  & \nodata & \nodata  \\
CH$_2$CHCN      & 1.5(16) & 9.4(-9)  & 4.1(15) & 2.6(-9)  \\     
HCOOCH$_3$      & 6.8(15) & 4.3(-9)  & 4.2(15) & 2.6(-9)  \\
 \noalign{\smallskip}
 \hline
\end{tabular}  \end{center}
\footnotesize
  \begin{flushleft}
\footnotesize     
  $^{\mathrm a}$ Values are taken from \citep{kua04}; $a(b) = a \times 10^b$.
  \end{flushleft} 
\end{table}
 

\clearpage

\clearpage

\clearpage

\clearpage

\end{document}